\documentclass[epj,referee]{svjour}

\usepackage{graphicx}
\usepackage{amsmath}
\usepackage{amssymb}
\usepackage{color}

\begin{document}

\title{Fractality of profit landscapes and validation of time series models for stock prices}
\author{Il Gu Yi\inst{1}, Gabjin Oh\inst{2}\thanks{Email: phecogjoh@chosun.ac.kr},
	and Beom Jun Kim\inst{1}\thanks{Email: beomjun@skku.edu}}
\institute{BK21 Physics Research Division and Department of Physics, Sungkyunkwan University, Suwon 440-746, Korea,
	\and Division of Business Administration, Chosun University, 501-759 Gwangju, Korea}
\date{Received: data / Revised version: date}

\abstract{
We apply a simple trading strategy for various time series of real and
artificial stock prices to understand the origin of fractality observed in
the resulting profit landscapes.  The strategy contains only two parameters $p$
and $q$, and the sell (buy) decision is made when the log return is larger
(smaller) than $p$ ($-q$). We discretize the unit square $(p,q) \in [0,1]
\times [0,1]$ into the $N\times N$ square grid and the profit $\Pi(p,q)$ is
calculated at the center of each cell.  We confirm the previous finding that
local maxima in profit landscapes are scattered in a fractal-like fashion: The
number $M$ of local maxima follows the power-law form $M \sim N^a$, but the
scaling exponent $a$ is found to differ for different time series.  
From
comparisons of real and artificial stock prices, we find that the fat-tailed
return distribution is closely related to the exponent $a \approx 1.6$
observed for real stock markets.
We suggest that the fractality of profit landscape characterized by
$a \approx 1.6$ can be a useful measure to validate time series model 
for stock prices. }
%

%
%

\authorrunning{I. G. Yi, {\it et. al.}}

\maketitle

\section{Introduction}
More and more physicists have been drawn to the research fields of
economics and finance in the last few decades~\cite{Stanleybook,Bouchardbook,Sinhabook}.
Those econophysicists brought new insights into such interdisciplinary area
and developed useful analytical and computational tools.
Many stylized facts in stock markets have been repeatedly discovered by researchers:
the heavy tail in the return distribution, rapidly decaying autocorrelation of returns,
the clustering and long-term memory of volatility, to list a few.

Recently, a geometric interpretation of the profit in stock markets has been
proposed, in which the profit landscape of a stock has been
constructed based on a simple virtual trading strategy~\cite{Gronlund}.
Although the used trading strategy is too simple to mimic behaviors of
real market traders, the simplicity has its own benefit: It allows us
to construct the profit landscape in low dimensions and thus making
the geometric analysis straightforward. More specifically, we use the
two-parameter strategies to compute the long-term profit of individual
stock in the real stock market.
Our study in this paper is an extension of Ref.~\cite{Gronlund}: We first show that
the observed fractality in Ref.~\cite{Gronlund} is a generic property of real stock markets,
by examining stock markets in two different countries (Unites States and South Korea).
We then pursue the origin of this observed fractality by applying the same
methodology to various time series, real and artificial. It is revealed that
the fractal structure in the profit landscapes in real markets is closely
related with the existence of fat tails in the return distribution.

The present paper is organized as follows: In Sec.~\ref{sec:methods}, we
present the datasets we use and also describe briefly how we generate
four different artificial time series to compare with real stock price
movements. Our two-parameter trading strategy is also explained.
Section~\ref{sec:results} is devoted for the presentation of
our results, which is followed by Sec.~\ref{sec:conc} for conclusion.


\section{Methods}
\label{sec:methods}

\subsection{Time Series}

In the present study, we employ a simple long-short trading strategy~\cite{Gronlund} and apply
it to the historical daily real stock price time series as well as to artificially
generated time series.  For real historic stock prices, we use 526 stocks in
South Korean (SK) for 10 years (Jan. 4, 1999 to Oct. 19, 2010), and 95 stocks
in United States (US)  for 21 years (Jan. 2, 1983 to Dec. 30, 2004). For SK
dataset, the total number of trading days is $T=2918$, while $T=5301$ for
US.  For artificially generated time series data, we use four different
models that are widely used in fiance:
the geometric Brownian motion (GBM), the fractional Brownian motion
(FBM), the symmetric L{\'e}vy $\alpha$-stable process (LP), and the Markov-Switching
Multifractal model (MSM). For each stochastic process, we generate 100 different time series
for the same time period $T = 2918$ as for the SK dataset. In Table~\ref{tab:time},
we list the properties of different time series used in the present study.

\begin{table*}
\center
\begin{tabular}{l|cccc|c}
\hline \hline
	&	GBM	&	FBM	&	LP	&	MSM	&	Real Markets	\\
\hline
long-term memory	&	$\times$	&	\footnotesize{$\bigcirc$}	&	$\times$	&	\footnotesize{$\bigcirc$}	&	\footnotesize{$\bigcirc$}	\\
heavy-tail distribution	&	$\times$	&	$\times$	&	\footnotesize{$\bigcirc$}	&	\footnotesize{$\bigcirc$}	&	\footnotesize{$\bigcirc$}	\\
\hline \hline
\end{tabular}
\caption{Stock price data from real markets have the long-term memory and the heavy-tailed return distribution.
Artificial financial time series produced by the geometric Brownian motion (GBM), the fractional Brownian motion (FBM),
the symmetric L{\'e}vy $\alpha$-stable process (LP), and the Markov-Switching Multifractal model (MSM) have different properties.
While FBM shares the long-term memory property with real markets, LP and MSM exhibit heavy-tails in return distribution
like real markets. See text for details.}
\label{tab:time}
\end{table*}

\begin{figure*}
\includegraphics[width=0.95\textwidth]{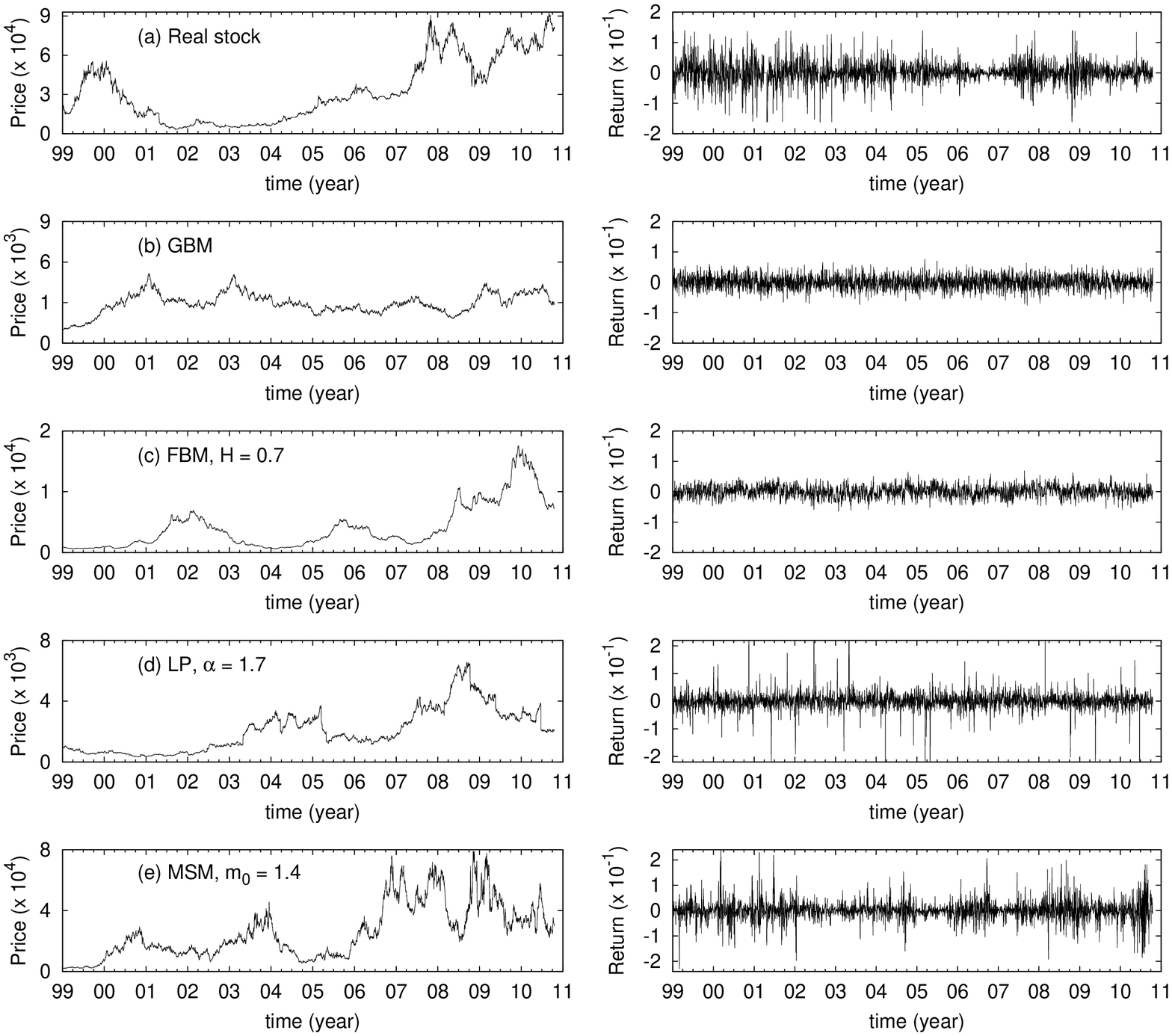}
\caption{The price movements (left column) and their corresponding log-returns (right column)
for (a) a real stock, (b) geometric Brownian motion (GBM),
(c) the fractional Brownian motion (FBM) with the Hurst exponent $H=0.7$,
(d) the symmetric L{\'e}vy $\alpha$-stable process (LP) with $\alpha=1.7$, and (e)
the Markov-Switching Multifractal model (MSM) with $m_0 = 1.4$.  For comparisons of behaviors,
see Table~\ref{tab:time}.
}
\label{fig:timeseries}
\end{figure*}

The GBM~\cite{Black,Hull} is based on the stochastic differential equation
\begin{equation}
\label{eq:GBM}
dS_{t} = S_{t} (\mu dt + \sigma d W_{t}),
\end{equation}
where $S_{t}$ is the price of stock and $W_{t}$ is the standard Brownian motion.
For the drift $\mu$ and the volatility $\sigma$ in Eq.~(\ref{eq:GBM}), we use the
values obtained from the SK dataset.

To mimic the existence of a long-term memory in real stock prices
the continuous time stochastic process FBM has been introduced~\cite{Mandelbrot}, in which
the covariance between the two stochastic processes $B_{t}^{(H)}$ and $B_{s}^{(H)}$ at time $t$ and $s$
satisfies

\begin{equation}
\label{eq:FBM}
\textrm{Cov} (B_{t}^{(H)}, B_{s}^{(H)}) = \frac{1}{2} \left[
|t|^{2H} + |s|^{2H} - |t-s|^{2H} \right].
\end{equation}
Here, the Hurst exponent $H \in (0, 1]$ plays an important role:
For $H=1/2$, FBM is the same as the standard Brownian motion, while
for $H>1/2$ ($H < 1/2$), the increments of its stochastic process
are positively (negatively) correlated. Accordingly, the long-term
positive memory exists for $H > 1/2$.
The long-term memory properties in various financial time series, such as
stock indices, foreign exchange rates, futures, and commodity, have been
found~\cite{Ding,Liu,Cont,Oh}.
In this work, we generate the process by using
\begin{equation}
\label{eq:FBS}
dS_{t} = S_{t} (\mu dt + \sigma dB_{t}^{(H)}),
\end{equation}
where $S_t$ is the stochastic variable for the stock price and $B_{t}^{(H)}$ is the
fractional Brownian motion [compare with Eq.~(\ref{eq:GBM})].
We use the package {\tt dvfBm} in \emph{R-project} to get the sample paths of FBM and
generate time series for different values of the Hurst exponent $H$.

We also generate the stochastic time series using LP~\cite{Sinhabook,Mantegna,Pantaleo,Schoutensbook,Fusaibook},
to imitate the heavy-tail distribution of return in real market (see Table~\ref{tab:time}).
We first produce the stochastic process $X_t$ by LP and then generate
the stock price $S_t$~\cite{Schoutensbook,Applebaumbook} via
\begin{equation}
S_{t} = S_{0} \exp(X_{t}).
\end{equation}
LP contains the tunable parameter $\alpha \in (0, 2]$ called
the stability index which characterizes the shape of distribution.
The distribution $P_\alpha(x)$ for the return $x$
converges to the Gaussian from as $\alpha \rightarrow 2$,
while for $\alpha < 2$ it asymptotically exhibits the power-law behavior
\begin{equation}
P_\alpha (x) \sim |x|^{-(1+\alpha)}, \
\end{equation}
for $|x| \rightarrow \infty$. Consequently, we can adjust the thickness of the tail part of
the distribution by changing $\alpha$.
We generate time series using LP for different values of $\alpha$.

MSM has been introduced to mimic the behaviors of real markets such as the
heavy-tailed return distribution, the clustering of volatility, and so
on~\cite{MFbook,Calvet2001,Calvet2004} (see Table~\ref{tab:time}).
It has become the one of the most well accepted models in the area of finance and econometrics.
To generate artificial stock prices $S_t$ based on MSM, we use
\begin{equation}
S_{t} = S_{t-1} \exp(R_{t}),
\end{equation}
where $R_t$ is the log-return written as
$R_t \equiv \sigma(M_{t}) \epsilon_{t}$.
The Gaussian random variable $\epsilon_t$ has zero mean and unit
variance, and the volatility $\sigma(M_t$) is constructed from
the time series $M_t$ in MSM~\cite{MFbook,Calvet2001,Calvet2004}.
In this work, we use the {\it binomial} MSM and use the parameter $m_0
\in (1,2)$ to generate different time series.

We generate artificial time series by using the above described four different
models (GBM, FBM, LP, and MSM) and compare them with real market behavior in
Fig.~\ref{fig:timeseries}, which displays the stock price and the log-return
time series in the left and the right columns, respectively. Qualitative
difference between artificial time series [(b)-(e)] and the real price (a) is
not easily seen. In contrast, the log-return shown in the right column of
Fig.~\ref{fig:timeseries} exhibits a clear difference.  In
Fig~\ref{fig:timeseries}(a) for the time series of the log-return for a real
company in SK, one can recognize well-known stylized facts: Fluctuation of
log-return around zero, called the volatility, has a broad range of sizes which
implies the heavy-tailed return distribution.  It is also seen that the
volatility is clustered in time so that there is an active period of time when
return fluctuates heavily and there is a calm period when return does not
deviate much from zero.  On the other hand, the two model time series GBM and
FBM in Fig~\ref{fig:timeseries} (b) and (c) show somewhat different behavior:
The size of fluctuation does not vary much, indicating the absence of the heavy
tail in return distribution.  In contrast, LP and MSM in
Fig.~\ref{fig:timeseries} (d) and (e) exhibit the larger volatility fluctuation
like the real market behavior in (a), indicating the heavy tail in return
distribution. The difference between LP and real market can be seen in terms of
the volatility clustering: In LP, the large volatility is not necessarily
accompanied by other large volatility [Fig.~\ref{fig:timeseries}(d)], whereas
for MSM [Fig.~\ref{fig:timeseries}(e)],  and real market
[Fig.~\ref{fig:timeseries}(a)], volatility is often clustered in time.

\begin{figure*}
\includegraphics[width=1.0\textwidth]{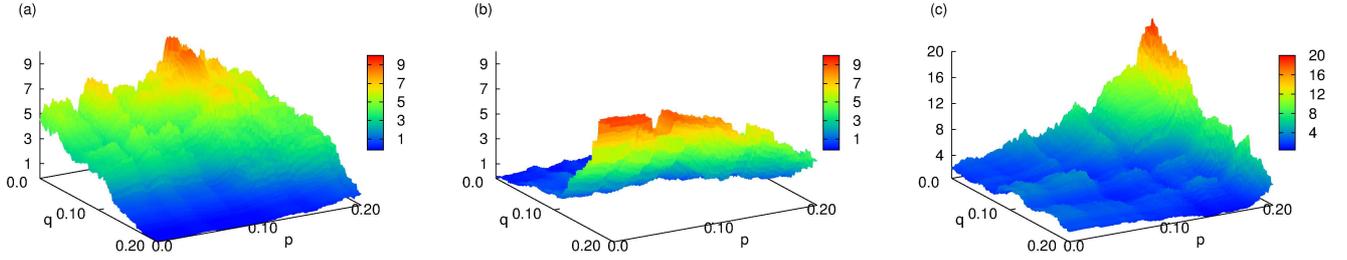}
\caption{(Color online) Profit landscapes for Samsung Electronics [(a) and (b)] and Hyundai Motors [(c)] in SK.
For (a) and (c) the strategy $S^{(1)}$ is used, and for (b) $S^{(2)}$ is used.
Although the landscapes look different from each other, they show qualitative similarity:
all are rough and have lots of local peaks and dips.
We have used $f_b = 0.5$, $f_s = 0.5$ and $d = 10$. See text for details.
}
\label{fig:landscape}
\end{figure*}

\begin{figure*}
\includegraphics[width=0.98\textwidth]{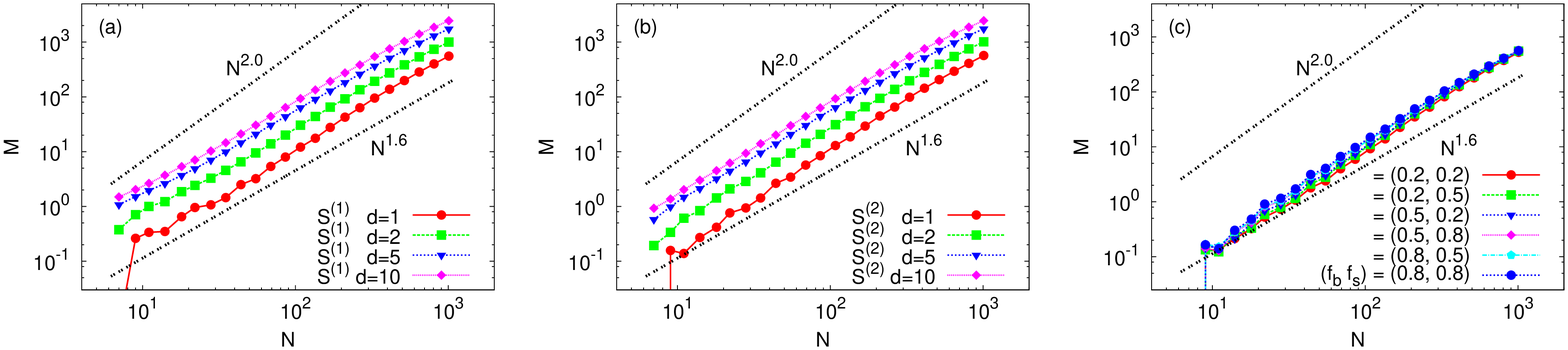}
\caption{(Color online) The number $M$ of local maxima, averaged over
526 stocks in SK, scales with the
resolution parameter $N$ following the power-law form $M \sim N^{a}$. In the
large $N$ regime, both strategies $S^{(1)}$ [(a) and (c)] and $S^{(2)}$ [(b)]
have $a \approx 1.6$ irrespective of the values of time lag $d$ [(a) and (b)] and
the values of $f_b$ and $f_s$ [(c)].  For (a) and (b), $f_b = f_s = 0.5$ is
used for $S^{(1)}$ and $S^{(2)}$, respectively, and for (c) $d=1$ is used. The
exponent $a \approx 1.6$, also observed for US market in the previous study~\cite{Gronlund},
is found to be a robust feature of real stocks in SK
and does not depend on details of trading strategy.}
\label{fig:fractal}
\end{figure*}

\subsection{Trading Strategy}
In our simple trading strategy (see Ref.~\cite{Gronlund} for more details),
the trade decision (buy or sell) is made when the log return
$R(t,t') \equiv \log [S_t/S_t']$ with $t > t'$ passes upper and lower boundaries:
Sell for $R(t,t') > p$, buy for $R(t,t') < -q$, and no trading if the log return
remains within boundaries $-q < R(t,t') < p$.
As in Ref.~\cite{Gronlund},
we further parameterize the strategy by introducing the time lag $d$ defined by
$d \equiv t - t'$, the fraction $f_s$ of holding shares of stock
when sell decision is made, and the fraction $f_b$ of cash in possession when buy.

We use the same notations as in Ref.~\cite{Gronlund} and call the above simple
strategy as $S^{(1)}$. In the inverse strategy $S^{(2)}$~\cite{Gronlund},
buy (sell) decision is made for $R(t,t') > p$ $[R(t,t') < -q]$.
It is interesting to see that $S^{(1)}$ is similar to the contrarian strategy
(or trend-opposing strategy) and that $S^{(2)}$ to the momentum strategy
(or trend-following strategy)~\cite{Gronlund}.
However, we emphasize that the similarity is only superficial:
The existence of trend has been strongly debated and in the area of econophysics
the well-known absence of temporal correlation of the return is clearly against
the existence of any trend in time scales longer than a couple of minutes~\cite{Gopikrishnan}.

The virtual trading strategy in this study is applied as follows:
\begin{enumerate}
	\item Start with one billion Korean Won ($m(1) = 10^{9}$) at $t=1$ as initial amount of
investment.~\footnote{The exchange rate between US Dollar and Korean Won (KRW) is about $1:1000$.}
	\item Pick a company $i$.
	\item For a given parameter pair ($p, q$), keep applying the trading strategy, $S^{(1)}$ or $S^{(2)}$,
until the last trading day $T$. For each trading, the amount of cash $m(t)$ and the number of stocks $n(t)$
change.
\end{enumerate}
At the end of the trading period $T$, we assess the total value of our virtual portfolio as
the sum of cash and the stock holdings:
$m(T) + n(T) S_T$.
The performance of the strategy is evaluated by the ratio
between the net profit and the initial investment, i.e.,
\begin{equation}
\Pi(p,q) \equiv \frac{m(T) + n(T)S_T - m(1)}{m(1)}.
\end{equation}
We constrain that the lower bound of cash is zero, i.e., $m(t) \geq 0$,
and that the number of shares is non-negative, i.e, $n(t) \geq 0$.
For the sake of simplicity, we also assume that the market price of stock is the same as the daily closing
price in the historic data, and the transaction commission fee $0.1 \%$ is taken into account for every
trading.

\section{Results}
\label{sec:results}

\subsection{Real time series}
We calculate the profit $\Pi^{i}(p,q)$ of each strategy for the $i$-th company
in order to study the fractality of profit landscape in stock markets.
The unit square $[0,1] \times [0,1]$ in $p$-$q$ plane is discretized by
the resolution parameter $N$ so that it contains $N^2$ small
squares of the size $(1/N) \times (1/N)$.
The shape of the profit landscape not only differs from company to company but also depends on the
details of strategy such as the values of $f_b$, $f_s$ and time lags $d$
as seen in  Fig.~\ref{fig:landscape} for two companies in SK: Samsung Electronics and Hyundai Motors.
We also find that the shape of the profit landscape and the locations of maxima and minima
totally change for different time interval
$t \in [1, T/2]$, $t \in [T/2+1, T]$, and $t \in [1, T]$~\cite{Gronlund}.
Even if the shape of the profit landscape looks quite different for various parameter and different
time interval, it is revealed that the nature of profit landscape does not change through Ref.~\cite{Gronlund}.

Once we get $\Pi^{i}(p,q)$ for a given resolution $N$, we count the number $M$
of local maxima, where the profit is larger than profits for its von Neumann neighborhood (the four
nearest cells surrounding the central cell).
Apparently, $M$ must be an increasing function of $N$, but how it depends on $N$ can be
an interesting question to pursue.
As is already observed for US stocks~\cite{Gronlund}, we again confirm that $M \sim N^{a}$ with
$a \approx 1.6$ as shown in Fig.~\ref{fig:fractal}, where $M$ is obtained from the average over
526 stocks in SK. Our finding of the same exponent $a \approx 1.6$ both for US and SK
indicates that the fractality of the profit landscape can be of a universal feature of different
stock markets.  We also check the robustness of the exponent $a \approx 1.6$
for different values of the parameters.  Figure~\ref{fig:fractal} (a) and (b) show that the same
scaling relation holds regardless of the specific strategy and different time lags $d$.
Figure~\ref{fig:fractal} (c) confirms $a \approx 1.6$ again for various parameter combinations of ($f_b$, $f_s$).
Although not shown here, $a \approx 1.6$ is again obtained when we use the Moore neighborhood (the eight cells
surrounding the central cell) instead
of the von Neumann neighbors, and when the size of parameter set is expanded to $(p, q) \in [0, 2] \times [0, 2]$.
These results seem to suggest that the fractal nature of the profit landscape is a genuine feature
of the real stock markets.

Although the trading strategies employed in this work might be too simple to be
used in reality, we believe that a complicated trading strategy can be
decomposed into a time-varying sophisticated mixture of simple strategies like $S^{(1)}$ and $S^{(2)}$.
Our observation of the robustness of the fractality suggests that more realistic strategies
beyond those used here must also have complicated profit landscape, which is expected 
to exhibit a fractal characteristic.

\subsection{Artificial time series}
\begin{figure}
\includegraphics[width=0.49\textwidth]{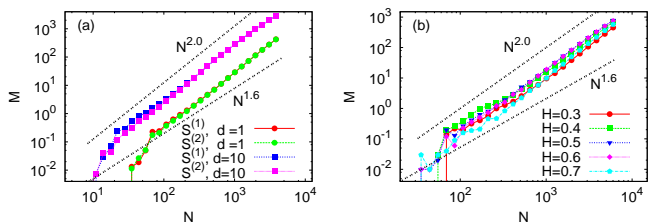}
\caption{(Color online) The scaling relation ($M$ versus $N$) yields $a \approx
2.0$ both for (a) GBM and (b) FBM.  In (b), the Hurst exponent $H$
characterizes the existence of the long-term memory, which hardly changes the
scaling exponent $a$ in the large $N$ regime. We have used $S^{(1)}$ with $f_b=f_s=0.5$.}
\label{fig:bm}
\end{figure}

We repeat the above procedure to construct profit landscapes for four different
artificial time series models GBM, FBM, LP, and MSM (see Sec.~\ref{sec:methods}).
Figure~\ref{fig:bm} shows $M$ versus $N$  for (a) GBM and (b) FBM.
Interestingly, we again find the scaling form $M \sim N^{a}$ but with $a \approx 2.0$ for both
GBM and FBM, which is significantly different from $a \approx 1.6$ observed in real markets.
It is to be noted that the value $a \approx 2.0$ indicates that the local maxima in the profit
landscapes for GBM and FBM are scattered like two-dimensional objects.
Also observed is that the scaling exponent $a$ does not change much as the Hurst exponent $H$ is
varied in FBM as shown in Fig.~\ref{fig:bm}(b). This is a particularly interesting result since FBM
with $H > 1/2$ is known to have the long-term memory effect like real markets. The difference
of $a$ between real markets and FBM implies that the nontrivial value $a \approx 1.6$ in real
markets may not originate from the long-term memory effect.

\begin{figure}
\includegraphics[width=0.49\textwidth]{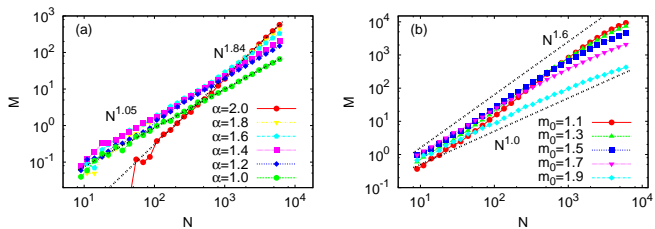}
\caption{(Color online) The scaling relation $M \sim N^a$ for (a) LP has the exponent $a$
which varies from 1.0 to 2.0 as the stability index $\alpha$ in LP changes. As $\alpha$ is
decreased from 2.0 (the Gaussian limit without heavy tail in return distribution), the
scaling exponent $a$ also decreases, indicating the close relation between the fractality
of the profit landscape and the tail thickness of the return distribution. (b) MSM yields
$a \approx 1.6$, reflecting the heavy-tailed return distribution. For both (a) and (b)
we have used $S^{(1)}$ with $f_b=f_s=0.5$.}
\label{fig:lpmsm}
\end{figure}

\begin{figure}
\centering
\includegraphics[width=0.40\textwidth]{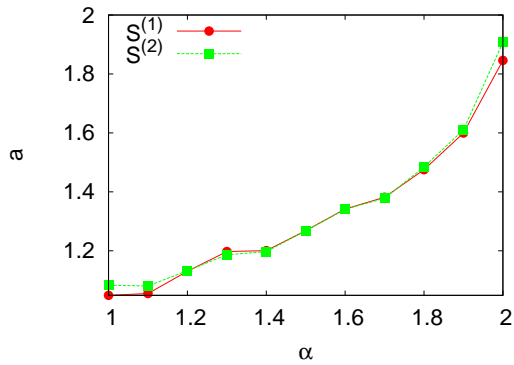}
\caption{(Color online) The scaling exponent $a$ versus the stability index $\alpha$ for LP.
It is shown that $a$ systematically decreases from the Gaussian value $(a \approx 2)$ toward $a \approx 1$ 
as $\alpha$ is decreased. The exponent $a$ hardly changes for different strategies $S^{(1)}$ and $S^{(2)}$.}
\label{fig:alphaA}
\end{figure}

We next construct profit landscapes for LP and MSM and measure the scaling exponents
in Fig.~\ref{fig:lpmsm}(a) and (b), respectively.
For LP, the stability index $\alpha$ controls the shape of the return distribution as described
in Sec.~\ref{sec:methods}. As $\alpha$ is decreased from the Gaussian limit $\alpha = 2$,
the tail part of the return distribution becomes thicker. In Fig.~\ref{fig:lpmsm}(a), it is clearly
shown that the scaling exponent $a$ which characterizes the fractality of the distribution of
the local maxima in the profit landscapes changes in a systematic way with $\alpha$. It starts from
the Gaussian limit value $a \approx 2$ [compare with Fig.~\ref{fig:bm}(a) for GBM] when
$\alpha = 2.0$ and systematically decreases toward $a \approx 1$, as displayed in 
Fig.~\ref{fig:alphaA}. Note also that $S^{(1)}$ and $S^{(2)}$ do not display any
significant difference in Fig.~\ref{fig:alphaA}.
Consequently, it is tempting
to conclude that the shape of the tail part of the return distribution (controlled by $\alpha$ in LP)
determines the scaling exponent $a$ in the profit landscape.
When $\alpha \approx 1.85$, we get $a \approx 1.6$ as for the real markets.
In Table~\ref{tab:time}, MSM is also shown to have the heavy tail in return distribution. From the study
of LP, one can expect that MSM should also lead to $a \approx 1.6$ as for LP. Figure~\ref{fig:lpmsm}(b)
displays that a proper choice of the parameter in MSM gives us $a \approx 1.6$. Another interesting
observation one can make from our MSM study is that $a \approx 1.6$ appears to be an maximum value
one can have, and the reason for this behavior needs to be sought in a future study.

In order to isolate the effects of the fat-tailed return distribution from the long-term memory effects,
we generate the time series of returns and fully shuffle them in time to produce a new stock price time series.
In this shuffling process, all existing long-time correlations are destroyed, but the probability distribution of
returns remains intact. We apply this method for real stocks and the MSM time series, and compute $M(N)$.
Although not shown here, we find that the scaling exponent $a \approx 1.6$ remains the same,
which again indicates that the fractal nature of the profit landscapes hardly depends on
the existence of long-term memory, as was already found for the LP.

\section{Conclusion}
\label{sec:conc}
In summary, we have investigated the origin of the fractality of the profit landscape in two different
real markets in comparison with four different models for generating artificial time series.
The number $M$ of local maxima has been shown to scale with the resolution parameter $N$ following
the form  $M \sim N^a$. Although the real markets exhibit $a \approx 1.6$, time series models with
the short-tailed return distributions (GBM and FBM) fail to produce this scaling exponent. In contrast, the two
different model time series (LP and MSM) have been shown to give us the scaling exponent consistent with
market value $a \approx 1.6$. In LP, a systematic tuning of the thickness of the tail part of
the return distribution has led to the main conclusion of the present work:
The origin of fractality in the profit landscape is the heavy-tail of the return distribution.
We suggest that the observed fractality of the profit landscape with the exponent $a \approx 1.6$
can play an important role in validation procedure of artificial time series models for
stock prices. In particular, one can tune the model parameters ($\alpha$ in LP and $m_0$ in  MSM, for example)
to make the time series compatible with the observed fractality in real markets.

\begin{acknowledgement}
B. J. K. (Grant No. 2011-0015731) and  G. O. (Grant No. 2012-0359)
acknowledge the support from the National Research
Foundation of Korea (NRF) grant funded by the Korea government (MEST).

\end{acknowledgement}

\end{document}